\documentclass[english]{ar-1col}
\usepackage[T1]{fontenc}
\usepackage[latin9]{inputenc}
\usepackage{babel}
\usepackage{amsmath}
\usepackage{graphicx}
\usepackage[unicode=true,pdfusetitle,
 bookmarks=true,bookmarksnumbered=false,bookmarksopen=false,
 breaklinks=false,pdfborder={0 0 1},backref=false,colorlinks=false]
 {hyperref}
\hypersetup{
 linktocpage}

\makeatletter
\usepackage{cite}

\makeatother

\begin{document}
\title{Beyond-Standard-Model Physics Associated with the Top Quark}

\author{Roberto Franceschini 
\\
\small{\it Universit\`a degli Studi and INFN Roma Tre, Via della Vasca Navale 84, I-00146, Roma}\\
email: \href{mailto:roberto.franceschini@uniroma3.it}{roberto.franceschini@uniroma3.it}
}

\begin{abstract}
We review scenarios of physics beyond the Standard Model in which
the top quark plays a special role. Models that aim at the stabilization
of the weak scale are presented together with the specific phenomenology
of partner states that are characteristic of this type of model. Further,
we present models of flavor in which the top quark is singled out as
a special flavor among the SM ones. The flavor and collider phenomenology
of these models is broadly presented. Finally, we discuss the possibility
that dark matter interacts preferably with the top quark flavor and
broadly present the dark matter phenomenology of these scenarios, as
well as collider and flavor signals.
\end{abstract}


\maketitle

\tableofcontents{}

\section{Introduction }

The top quark is a singular object amidst the fermions of the Standard
Model as it is the heaviest among them. This entails several peculiar
properties: in the domain of QCD it stands out as it is the only quark
never to be observed into a hadron, as its decay is much faster than
the hadron formation time; after the discovery of the Higgs boson,
and for all the time before in which the Higgs mechanism has dominated
the landscape of model building for the electroweak sector of the
SM, the top quark stands out as the only one with a ``normal'' size
coupling with the Higgs boson. This latter property has made the top
quark very interesting both for the question about the origin of the
structure of flavor in the SM and for the origin of the electroweak
scale itself. The special interest about top flavor has to do with
its strong preference to decay into bottom quarks, i.e not involving
other flavor families, which in the $CKM$ picture results in $V_{tb}=1$
up to small corrections, and its large mass, which can possibly act
as a magnifier of the effects of physics beyond the Higgs boson as
origin of flavor. For electroweak physics the top quark plays a crucial
role in that it affects the properties of the Higgs boson, and by
the Higgs mechanism for weak bosons mass generation, also in the physics
of weak gauge bosons: its effect can be seen in their masses and decay
rates, which are sensitive to the strength of the top quark gauge
and Yukawa couplings and to its mass. Deviations of these properties
from the SM predictions can be signs of new physics related to the
top quark. While the importance of the top quark can be appreciated
already from these general facts, the detailed role played by the
top quark can be better understood going closer to explicit new physics
models, which will pace the exposition of the greatest part of the
following material.

In sections~\ref{subsec:Supersymmetry} and \ref{subsec:Composite-and-pNGB/Little}
we discuss models in which the top quark plays a special role for
the origin of the electroweak symmetry. The discussion is further
extended in section~\ref{sec:EFT} in a more model independent direction
using a flavor-conserving effective field theory of the top quark
sector, which also allow to discuss prospects for top quark physics
at future colliders. In section \ref{subsec:Top-quark-and-flavor}
we attack a different problem, that of the origin of the flavors of
the SM. In section~\ref{subsec:Flavored-dark-matter} we extend the
discussion to the possibility that SM flavor plays a part in the stabilization
of the dark matter in a way that makes the dark matter interact preferably
with the top quark flavor and discuss the phenomenology of dark matter
in these scenarios. Finally in section~\ref{sec:Conclusions} we
offer some conclusions. 

Being the subjects list rather large, the discussion is necessarily
kept free from some details, which are available in the provided references.
This review is conceived so that it can also be useful for younger
graduate students seeking an high-level introduction to the subject(s)
discussed. Hopefully the readers can start here their own exploration
on topics that would otherwise require to go through a large stack
of literature. References are kept to a minimum of key works as to
encourage the reader to actually study these selected works.

\section{Top quark and BSM related to the Higgs boson and the origin of the
weak scale}

\subsection{Supersymmetry\label{subsec:Supersymmetry}}

Supersymmetry has been proposed as a space-time symmetry involving
fermionic generators. Unlike in gauge symmetries, this makes possible
to involve spin and momentum in the definition of the symmetry algebra,
which, up to violations of the symmetry itself, would require interactions
and masses of bosonic and fermionic particles to be tightly related.
One such relation would require the electron to be accompanied by
exactly mass degenerate states of spin-0, pretty much the same as
Lorentz symmetry of space-time built-in the Dirac equation implies
the existence of exactly mass degenerate anti-particles of the electron.
The absence of any evidence in experiments for spin-0 electron-like
state motivates to consider supersymmetry as an approximate symmetry,
broken at some unknown scale so that all the supersymmetric partners
of the SM states are pushed beyond the mass scale presently probed
by experiments. 

The mechanism for supersymmetry breaking is a subject for model building,
which is outside of the scope of this review. For our purpose it is
key to recall that the supersymmetry breaking top quark sector has
the rather model-independent tendency to determine the Higgs bosons
mass and quartic coupling, thus leading to the identification of the
supersymmetric top scalar quark, most often called ``stop squark'',
as the main player setting the Higgs boson potential. In the Minimal
Supersymmetric Standard Model (see \cite{Martin:1998ve} for an extensive
review) this is represented by equations for the constraints on the
minimization of the Higgs boson potential 
\begin{align*}
m_{Z}^{2} & =\frac{\left|m_{H_{d}}^{2}-m_{H_{u}}^{2}\right|}{\sqrt{1-\sin^{2}(2\beta)}}-m_{H_{u}}^{2}-m_{H_{d}}^{2}-2\left|\mu\right|^{2}\,,\\
\sin(2\beta) & =\frac{2b}{m_{H_{u}}^{2}+m_{H_{d}}^{2}+2\left|\mu\right|^{2}}\,,
\end{align*}
coupled with the 1-loop effect of the top quark and top squark on
the bilinear terms of the 2 Higgs doublets $H_{u}$ and $H_{d}$ .
In particular, for the Higgs doublet $H_{u}$ that interacts with
up-type quarks, hence feels the top quark sector, the RGE equations
is
\begin{align}
\frac{d}{d\log Q}m_{H_{u}}^{2} & =3X_{t}-6g_{2}\left|M_{2}\right|^{2}-\frac{6}{5}g_{1}^{2}\left|M_{1}\right|^{2}+\frac{3}{5}g_{1}^{2}S\,,\label{eq:dmHu-dt}
\end{align}
 where $X_{t}=2\left|y_{t}\right|^{2}\left(m_{H_{u}}^{2}+m_{Q_{3}}^{2}+m_{\bar{u}_{3}}^{2}\right)+2\left|a_{t}\right|^{2}$,
$M_{1,2}$ are the U(1) and SU(2) gaugino mass terms, and $S=Tr[Y_{j}m_{\phi_{j}}^{2}]$.

These equations naturally lead to possibility that the supersymmetry
breaking stop masses $m_{Q_{3}}^{2}$ and $m_{\bar{u}_{3}}^{2}$ or
a large A-term $\left|a_{t}\right|$ might induce a large $X_{t}$,
which in turn drives $m_{H_{u}}^{2}<0$ as $\log Q$ diminishes from
some high-scale down to the weak scale. This possibility has made
the role of stop squarks a very central one in supersymmetric models.
In essence, the supersymmetric partner of the top quark is responsible
for breaking the electroweak symmetry, by making $m_{H_{u}}^{2}<0$
hence making the Higgs boson potential unstable at the origin of the
$H_{d},H_{u}$ fields space, and setting the value of the masses that
set the weak scale, e.g. $m_{Z}$ from the above equation or the mass
of the Higgs boson that receives the above mentioned large radiative
corrections from the stop squark.

As a matter of fact, once the Higgs boson was discovered and its mass
was known, a number of works tried to determine the impact of this
measurement on the properties of the stop squark (e.g. see Ref.~\cite{Hall:2011rt}
for the MSSM and some extensions). In turn, the necessity for peculiar
supersymmetry breaking to accommodate the Higgs mass has spurred investigations
on the possible supersymmetry breaking models that can lead to such
peculiar stop squarks (see e.g. \cite{Larsen:2012ys,1204.5977v2,Craig:2012lq,Craig:2012dq,1112.6261v2,1203.0572v1}
for some examples of supersymmetry breaking models emerged or re-emerged
to address the null searches of supersymmetry and the Higgs discovery).

\subsubsection{Phenomenology}

The phenomenology of the supersymmetric partners of the top quark
is largely dictated by one feature of the supersymmetric models: the
existence of a conserved quantum number that distinguishes SM states
from their supersymmetric partners. The standard choice for such quantity
is called $R$-parity, a $Z_{2}$ symmetry under which all SM states
are even and all partners states are odd. The conservation of this
symmetry implies that partners states can appear in interaction vertexes
only in even number, e.g. one SM states can interact with two supersymmetric
states and it is not possible for a single supersymmetric state to
interact with a pair of SM states. For particle colliders this implies
that the lowest order process to produce supersymmetric states in
collisions is 
\[
SM\,SM\to SUSY\:SUSY,
\]
and the decay of supersymmetric particles to any number of SM states
is forbidden unless there is at least one supersymmetric particle
(or an odd number of them), e.g. 
\[
SUSY\to SUSY\,SM\,.
\]
When $R$-parity is exact a most copious production mechanism for
stop squarks at the LHC is
\begin{equation}
gg\to\tilde{t}_{i}\tilde{t}_{j}^{*},\label{eq:stopprodution}
\end{equation}
where we denoted $\tilde{t}_{k}$ for $k=1,2$ the two stop squarks
mass eigenstates~\footnote{The definition of mass eigenstate as ``stops'' assumes that flavor
labels we give in the SM are the same for the partners states. It
must be stressed that the fate of flavor in the supersymmetric partners
sector is largely model dependent and it is possible to use flavor
mixing to change the phenomenology of stop squarks, see e.g. \cite{1302.7232v2}.
See \cite{Martin:1998ve} for more details on the gauge and flavor
structure of the squark sector.}. Other production mechanisms are possible, e.g. in decays of supersymmetric
partners heavier than the stops or via production of stops in association
with other supersymmetric states.

Once produced, the stop squark can decay in a number of possible channels,
depending on which supersymmetric states are lighter than the state
$\tilde{t}_{k}$ at hand. Most studied 2-body decay modes are
\begin{equation}
\tilde{t}\to t\chi^{0},\tilde{t}\to b\chi^{+}\,,\label{eq:stop-2b-decay}
\end{equation}
which feature fermions $\chi$ that are mixtures of supersymmetric
partners of gauge bosons of the electroweak interactions and of the
Higgs bosons of the model. The motivation for the prevalence of these
decay modes is that, by the rules of unbroken supersymmetry, these
decays are mediated by couplings given by gauge and Yukawa couplings
of the SM, hence they are pretty much impossible to switch off unless
$m_{\tilde{t}}-m_{\chi}<0$. As a matter of fact the quantity $m_{\tilde{t}}-m_{\chi}$
plays a major role in determining the stop phenomenology. When $m_{\tilde{t}}-m_{\chi}\to0$
it becomes necessary to consider multi-body processes are also possible
and may be phenomenologically relevant, e.g.
\begin{equation}
\tilde{t}\to bW^{+}\chi^{0},\,\tilde{t}\to b\bar{f}f^{'}\chi^{0},\label{eq:stop-3b-decay}
\end{equation}
as well as possible flavor violating decays that may be induced at
loop level, such as 
\begin{equation}
\tilde{t}\to c\chi^{0}\,.\label{eq:stop-flv-decay}
\end{equation}
In the above discussion the particle $\chi^{0}$ is considered as
the lightest supersymmetric state (LSP), so that, by the conservation
of $R$-parity, it is absolutely stable. As $\chi^{0}$ is not electrically
charged and it is color neutral, pretty much like neutrinos it does
not leave directly observables traces in detectors. For this reason
the presence of $\chi^{0}$ can be detected only as momentum missing
in the overall momentum conservation in each collision. As we cannot
reliably measure the fractions of the longitudinal momentum of the
colliding protons taken by the partons initiating the production of
stops, e.g. the gluons entering in eq.(\ref{eq:stopprodution}), and
the fraction taken by the rest of the partons, the longitudinal momentum
conservation is usually not exploited in hadron colliders, therefore
the presence of $\chi^{0}$ is usually sought for as missing transverse
momentum, most often (mis)named missing transverse energy $mET$. 

Being an electrically neutral stable particle charged only under supersymmetric
Yukawa and electroweak gauge interactions, $\chi^{0}$ qualifies as
perfect candidate for a WIMP Dark Matter. The possibility to have
a Dark Matter candidate stemming out of supersymmetry has given formidable
motivation to pursue this scenario for the past decades. So much so,
that missing transverse energy searches have becomes synonymous of
searches for supersymmetry. It must be said, however, that the null
searches of supersymmetric particles, as well as WIMP Dark Matter
in the mass range suitable for $\chi^{0}$\cite{Aprile:2018aa}, has
put this idea under great pressure lately~\cite{Craig:2013yu,Feng:2013ys}. 

Given these experimental results, and the vast range of possible models
for supersymmetry breaking, it must be recalled that in general it
is possible to have other states than $\chi^{0}$ as lightest supersymmetric
particles. For instance the supersymmetric partner of a neutrino or
even top sector squarks. The latter leads to peculiar phenomena due
to the formation of hadrons containing supersymmetric states\cite{Fairbairn:2007cs}\cite{Sarid:2000lr},
but these models typically suffer from quite stringent limits \cite{ATLAS:2016onr,ATLAS:2019jcm,Kraan:2005ji}.
Therefore the majority of the searches for supersymmetric states in
the top quark sector are carried out in the $\chi^{0}$ LSP setting.

Wholly alternative phenomenological scenarios for supersymmetric top
quark partners are possible and are actively pursued in experimental
searches. The main possible alternative has to do with the non-conservation
of $R$-parity~\cite{Barbier:2005rr}. With broken $R$-parity all
supersymmetric particles can in principle be produced singly and can
decay into just SM states, e.g.
\[
SM\,SM\to SUSY\:\text{and }SUSY\to SM\,SM\,,
\]
are now possible processes. In this situation there is no longer an
absolutely stable weak scale particle to purse the idea of Dark Matter
as a WIMP\footnote{Alternative DM candidates can be found in these models, see e.g. \cite{Takayama:2000hc}
for a possible gravitino dark matter scenarios and issues related
to this possibility.} and the phenomenology of supersymmetric states linked to the top
quark is now greatly different from the picture given above~\cite{Franceschini:2015xy}.
For instance $R$-parity violating couplings, still respecting the
full gauge symmetry of the SM, allow, among other possibilities, the
decays 
\[
\tilde{t}\to bs\text{ or }\tilde{t}\to\ell d\,.
\]
As the final states of stop decay can now be made entirely of SM particles
it is possible to detect stop squarks as resonances, a very powerful
signature, that is not possible to pursue when $\chi^{0}$ is forced
to appear among the decay products. Furthermore these decays, being
mediated by $R$-parity breaking couplings, that need to be small
for a number of constraints~\cite{Barbier:2005rr}, can lead to meta-stable
supersymmetric states, which can live measurable lengths in experiments. 

\subsubsection{Experimental searches}

In a detailed model it is possible to derive very specific signals
from top sector supersymmetric partners, including both signatures
at collider experiments and as well as low energy precision ones.
The latter, however, turn out to be usually very much dependent on
the model considered for low energy precision experiments~\cite{Gabbiani:1996bv}.
A similar issue exists with early universe physics, on top of the
signals being quite difficult to detect. For this reason collider
searches are the prime way to search for top sector supersymmetric
partners.

Before listing relevant searches it is necessary to clarify a point
on their scope. The above searches are sensitive in principle to \emph{any}
sign of new physics related to the top quark sector involving mET
or some kind of pair produced resonances. Although the search is optimized
for supersymmetric partners, it can indeed be used to set bounds on
other models. The interested reader can refer for instance to Ref.~\cite{Kraml:2016ad}
for an interpretation of the ``supersymmetry searches'' in the context
of fermionic top partners to be discussed in later Section~\ref{subsec:Phenomenology}. 

The searches for top sector supersymmetric partners can be divided
into two main categories:
\begin{itemize}
\item searches in large momentum transfer signals, which feature detector
objects (jets, leptons, photons, ...) with energy and transverse momentum
greater than the typical SM events;
\item searches in low momentum transfer signal, in which the detector objects
arising from top sector supersymmetric partners production are not
very different from that of typical SM events.
\end{itemize}
The large momentum transfer ones are ``classic'' searches for new
physics, and were envisaged already at the time of design of the experiments~\cite{Bayatian:942733,ATLAS:1999vwa}.
Currently these searches can probe supersymmetric top partners up
to a mass around 1.2~TeV, although not in full generality. Indeed
it is quite hard to probe in full generality even a model as ``minimal''
as one having the full freedom to vary the branching ratios of decays
eqs.(\ref{eq:stop-2b-decay})-(\ref{eq:stop-flv-decay}). For a complete
assessment is then necessary to test very accurately a large number
of searches at once, often relying on a ``phenomenological'' incarnation
of a sufficiently general supersymmetric model, as studied for instance
in Ref.~\cite{ATLAS-Collaboration:2015it}. The interpretation of
these results is quite difficult, as many constraints on the model
are imposed at once, e.g. the top partners states are required to
``fix'' the mass of the SM Higgs boson to its measured value by
the dynamics of radiative corrections embodied in eq.(\ref{eq:dmHu-dt}).
This requirement, while being a sensible one in the context of the
specific model, can significantly alter the conclusion of that study.
Therefore it remains difficult to answer questions as simple as finding
the lightest not excluded values of the mass of stop-like top partners~\footnote{One possible answer in the context of \cite{ATLAS-Collaboration:2015it}
is offered in the supplementary material of that analysis\cite{pMSSM:stop:2015}. }.

\begin{figure}
\begin{centering}
\includegraphics[width=1\linewidth]{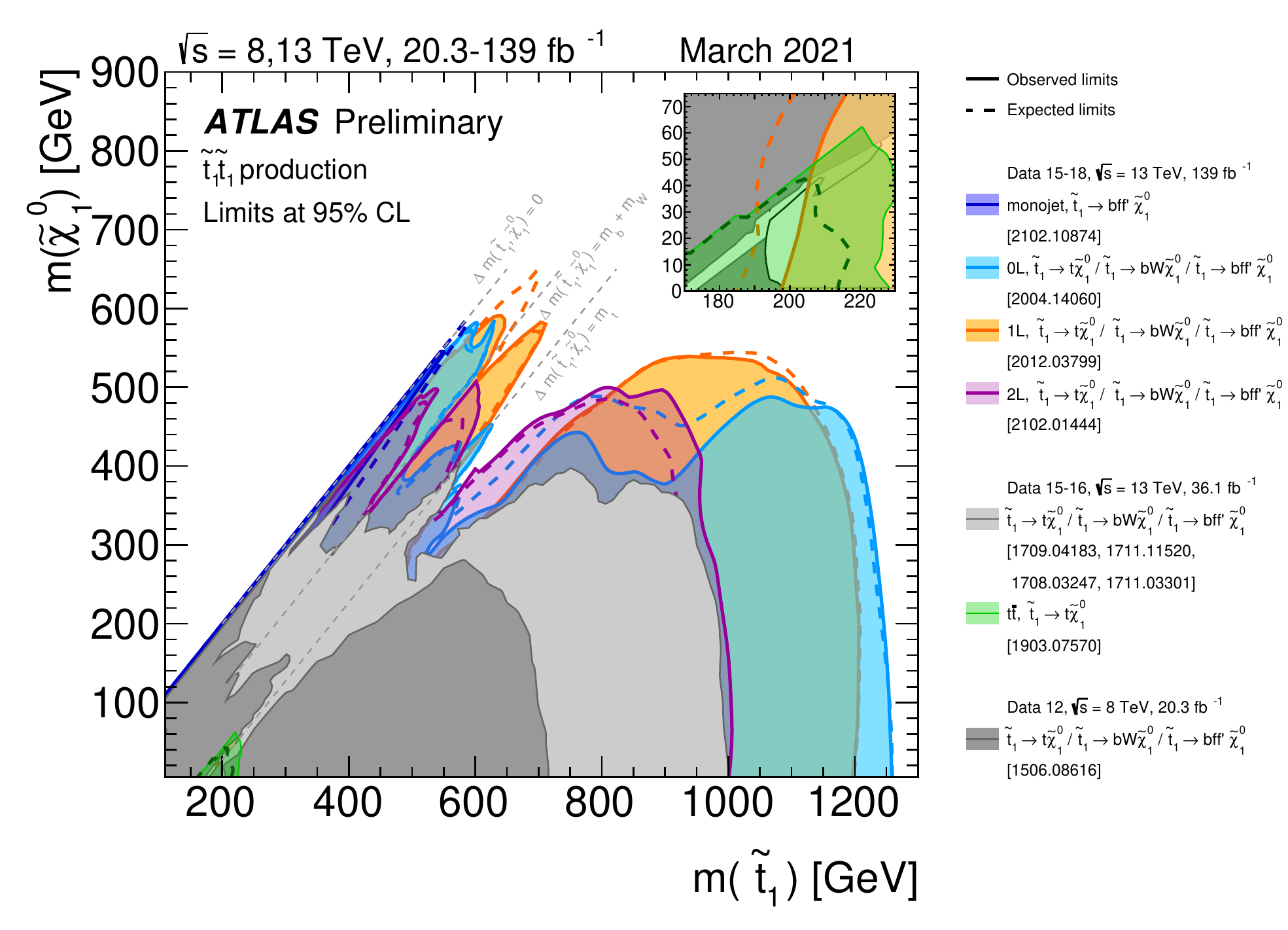}
\par\end{centering}
\caption{\label{fig:Searches-for-stops}Searches for top sector supersymmetric
partners in the Stop-LSP mass plane.}

\end{figure}

Further difficulties can arise and make nearly impossible to probe
experimentally supersymmetric top partners, e.g when special kinematical
configurations become the typical configuration of top partners decay
products. In these cases the search in low momentum transfer signatures
can help. Indeed, these searches have been developed to overcome the
difficulty that arise in the limit $m_{\tilde{t}}-m_{\chi}\to0$.
The shortcomings of the large momentum transfer searches can be clearly
seen in Figure~\ref{fig:Searches-for-stops}, as the excluded stop
mass for large $m_{\tilde{t}}-m_{\chi}$ is much larger than for small
values of this mass difference. In addition, when the stop-LSP mass
gap is small and the stop becomes lighter, its production and decay
cannot be reliably distinguished from other SM processes, e.g. the
SM top quark production. This observation motivates a zoom inset in
the figure to display how these peculiar cases are covered. The most
useful strategies to attack these difficult signatures have turned
out to be the studies of angular observables and fiducial rates of
top-like final states \cite{Godbole:2015it,Czakon:2014qf,Han:2012ve,1911.11244v1,Englert:2019ab}.
Especially in angular observables there are modest, but persistent
disagreement between the measurements in the top quark sample~\cite{ATLAS-Collaboration:2019ad}
and theoretical predictions. These disagreement are also accompanied
by other disagreements of small entity, but persisting from Run1 LHC
through Run2, in the kinematics of the reconstructed top quarks e.g.
in Refs.~\cite{2108.02803v2,ATLAS-Collaboration:2017ae}. The possibility
to see effects of BSM related to the top quark and the precision in
measurements afforded by the LHC and the HL-LHC has motivated the
great improvement of predictions for top quark SM observables, e.g.
\cite{Jezo:2016oj} for a seamless description of fixed NLO and PS
calculations of top quark resonant and non-resonant rates, \cite{Behring:2019aa,Bernreuther:2015fv}
for specific NNLO and EW corrections to the BSM sensitive rates and
more in general drawing attention on possibly BSM-sensitive high energy
top quarks (see e.g. \cite{2101.06068v1}) and other production modes
which may be of interest for both SM studies and BSM searches (see
e.g. \cite{2108.01089v1,Bevilacqua:2022ab}).

The searches mentioned above, though motivated and sometimes optimized
on supersymmetry searches, are rather general. Thus it is important
to stress that the observation of an excess in one of these ``supersymmetry
searches'' would not at all prove the supersymmetric nature of the
discovered state. A reliable statement on the supersymmetric nature
of the newly discovered object would require several measurements.
For some proposal at the LHC the interested reader can look for instance
at \cite{Kribs:2011ly}. In general it is believed that a machine
cleaner than a hadron collider, e.g. an $e^{+}e^{-}$ collider, capable
of producing the new particle would be needed to truly confer it the
status of ``supersymmetric partner'' state of some SM state.

At the time of writing there are no statistically significant and
convincing signs of new physics in searches for new physics, the searches
for supersymmetric top partners being no exception. Despite the absence
of signals for top sector supersymmetric partners these are still
believed to one our best chances to find new physics. Looking at the
glass as ``half full'' one could even argue that in the minimal
model of supersymmetry the relatively large observed Higgs boson mass
requires large loop level corrections from contributions of the kind
of eq.(\ref{eq:dmHu-dt}). These large loop corrections point towards
a stop squarks mass scale at the TeV or larger, thus perfectly compatible
with the present limits and possibly awaiting us for a next discovery
at one of the next updates of the searches as more data is collected
at the LHC. 

As the mass scale of top quark supersymmetric partners is not entirely
fixed it often considered that these particles may be too heavy for
the LHC to discover them. Therefore the discovery reach for these
particles is often considered in the evaluation of the physics case
of future particle accelerators. Projections for a 100~TeV $pp$
collider \cite{Mangano:2651294,Benedikt:2653674} usually cover a
mass range 5-8 times larger than what can be probed at the LHC, while
the expectation for a high energy lepton collider, such as multi-TeV
muon collider\cite{Black:2022ab,2203.07261v1,2203.07224v1,2203.07964v1,2203.08033v1},
is to probe the existence of top partners up to the kinematic limits
at $\sqrt{s}/2$.

\subsection{Composite and pNGB/Little Higgs\label{subsec:Composite-and-pNGB/Little}}

\subsubsection{Models }

New physics associated to the top quark sector has been motivated
also from a series of model building activities aimed at explaining
the origin of the electroweak scale through the Goldstone boson nature
of the agent of its breaking, resulting in theories of the Higgs boson
as a pseudo Nambu-Goldstone boson. From a low energy effective point
of view these theories can be put in the language of a composite Higgs
boson, whose lightness compared to its scale of compositeness is justified
by its goldstonian nature. Models built in this family are reviewed
in Refs.~\cite{Panico:2015lr,Perelstein:2006qy,Perelstein:2007fk,Contino:2010kx}
and they all share the need to enlarge the symmetries of the SM by
a new global symmetry, that is broken at some scale above the TeV
to a smaller symmetry, with the associated Nambu-Goldstone bosons,
which will host the yet smaller symmetry group of the SM at even lower
energies. The minimal model of this type \cite{Agashe:2004rs} that
is able to pass bounds from electroweak precision tests including
$Zb\bar{b}$ couplings assumes an $SO(5)$ global symmetry, broken
to $SO(4)\simeq SU(2)\times SU(2)$ which contain the weak interactions
gauged $SU(2)$.

The enlargement of the symmetry of the SM motivates appearance of
matter representations in multiplets that are necessarily larger than
the usual doublets and singlets of the SM. In particular, in order
to obtain Yukawa interactions the constructions of pNGB and little
Higgs model converges in the existence of ``partner'' states for
the top quark, the bottom quark and in principle for all the fermions
of the SM. The precise phenomenological manifestation of the ``partner''
states is highly model dependent, as it depends on the choice the
new global symmetry group that one has in building this type of models,
the representation of this symmetry group that one chooses for the
new matter and the imagined mechanism to originate the SM fermion
masses at the most microscopic level. 

One possible limitation to the model building choices may comes from
the requirement of not introducing large deviations in well known
couplings, e.g. the $Zbb$ couplings~\cite{Agashe:2006aa}, still
a large set of possibilities exists. For this review we focus on a
unifying feature of many models, that is the presence of ``partner''
states directly connected to the SM top quark sector via Yukawa and
gauge interactions with relatively universal decay patterns~\cite{De-Simone:2013to,Buchkremer:2013kl,Perelstein:2004aa},
although other decay modes and more ``exotic'' partners may exist
including possible couplings to scalar states accompanying the Higgs
boson in some models~\cite{Kearney:2013kx,Serra:2015ai,Cacciapaglia:2019ab}.

\subsubsection{Phenomenology\label{subsec:Phenomenology}}

At the core of the experimental tests of the idea of fermion top partners
lies the assumption that the main interaction leading to the decay
of these top partners into SM states is the Yukawa of the top quark,
in which the Higgs boson or longitudinal components of the weak gauge
bosons appear. For this reason the large majority of the searches
are presented in terms of exclusions for branching fractions of the
top partners states into the following pairs of SM states 
\[
T\to tZ,th,Wb,\,
\]
where $T$ is a charge 2/3 top partner and 
\[
B\to bZ,bh,Wt,
\]
where $B$ is a charge -1/3 partner of the bottom quark, whose existence
is consequence of the SU(2) weak isospin symmetry that must hold in
the theory that supersedes the SM at high energies. In models with
a symmetry larger than SU(2), e.g. \cite{Agashe:2006aa}\cite{Agashe:2004rs},
it is typical to have further partners states that appear as necessary
to furnish full representations of the larger symmetry. A much studied
case is the state of charge 5/3 that leads to a very characteristic
decay
\[
X_{5/3}\to W^{+}t\,,
\]
which in turn gives a characteristic same-sign di-lepton signal~\cite{Contino:2008aa}.
For little Higgs models the appearance of this type of exotic partners
requires the formulation of somewhat more involved models, but it
is definitively a possibility\cite{Kearney:2013kx,Kearney:2013fj}.

\subsubsection{Experimental searches at colliders}

Experimental searches for new states are carried out at the LHC exploiting
the color charge of the top partners in processes such as
\[
gg\to TT\,,
\]
that are analogous to previous processes for supersymmetric partners
and depend only on the QCD charge of $T$. Unlike for supersymmetric
partners, for which the conservation of R-parity plays a crucial role,
the single production of top partners 
\[
gq\to q'Tb\,,
\]
is possible in the most minimal models and can in principle lead to
a deeper understanding of the BSM physics, as this process involves
directly new physics couplings for the production of the top partners
state~\cite{Mrazek:2010ve}. For instance the rate of single production
of top partners states can be a discriminant with respect to so-called
``vector-like'' quarks, whose couplings are not dictated by Goldstone
property of the Higgs (see Ref.~\cite{Matsedonskyi:2016aa} for a
more in-depth discussion). 

A great difference in the search for the partners discussed in this
section is that they can in principle give rise to resonant signals,
e.g. in the invariant mass of an hadronic top and one hadronic Higgs
boson in the decay $T\to th$ and other signals discussed for instance
in the search of Ref.\cite{ATLAS:2018uky}. 

Another consequence of the top partner decaying in purely SM final
states is that even the ``heavy'' SM particles, such as $t$, $Z$,
$W$, $h$, are produced with significant boost in the majority of
the events. This motivates the use of special experimental techniques
for the identification of those detector objects~\cite{doi:10.1146/annurev-nucl-102419-055402}
as for instance in the search of Ref.\cite{ATLAS-Collaboration:2022ab}.

\begin{figure}
\begin{centering}
\includegraphics[clip,width=0.99\textwidth]{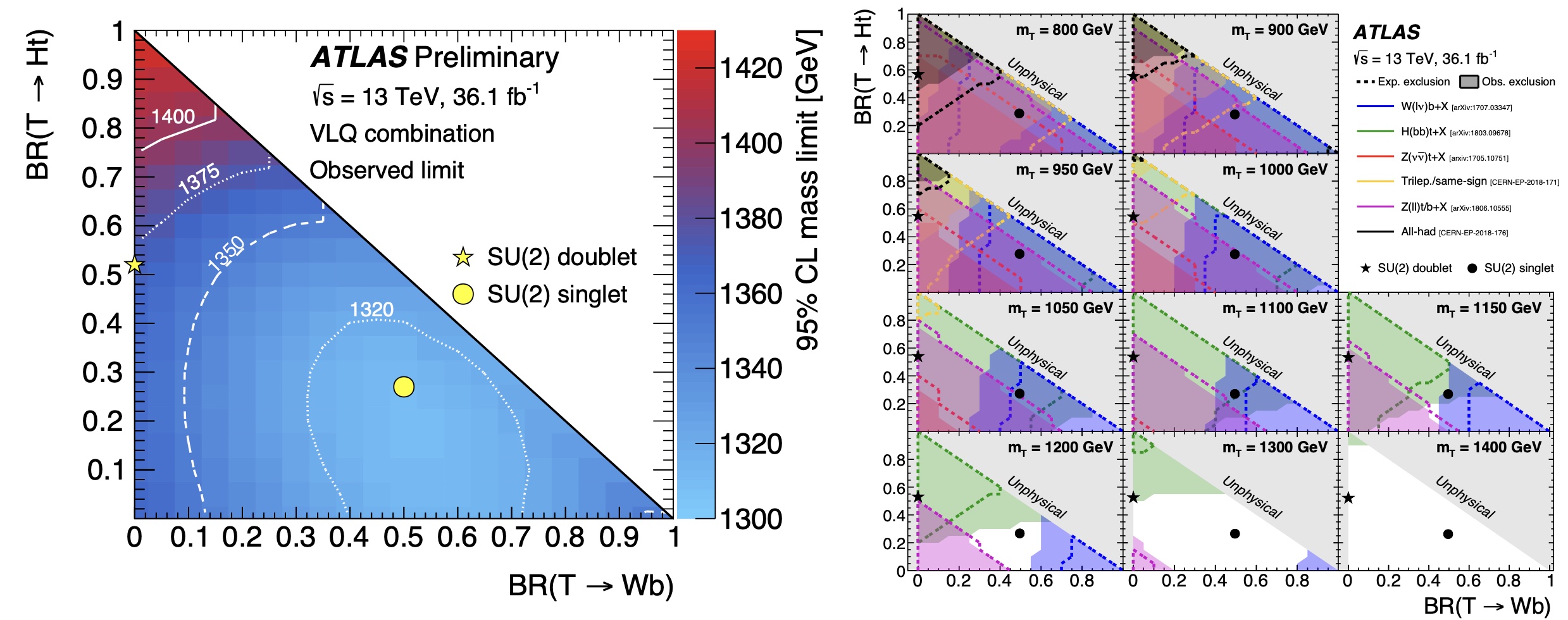}
\par\end{centering}
\caption{\label{fig:Searches-for-fermionic-partners}Searches for top fermionic
partners \cite{ATLAS-Collaboration:2018ae,ATLASexoticsSummary} in
the plane $BR(T\to Ht)$ vs $BR(T\to Wb)$ with the constraint $B(bW)+B(th)+B(tZ)=1$.
For reference, some model-dependent choices of the branching ratios
introduced in Ref.\cite{AguilartextendashSaavedra:2009aa} are shown.}
\end{figure}

The search strategies mentioned above are combined by the experimental
collaborations, that present results in a plane with axes spanning
the possible values of two decays, e.g. if figure~\ref{fig:Searches-for-fermionic-partners}
an example is shown for $T\to Ht$ and $T\to Wb$. The underlying
assumption of this presentation of the results is that the top partner
does not decay in any BSM states, hence the branching ratio of $T\to Zt$
is determined by the two branching rations displayed. The right panel
of the same figure shows how the different searches have different
sensitivity to each decay mode and can be patched together to better
exclude top partners of a given mass. For more exotic signals from
$X_{5/3}$ searches are carried out as well, e.g. in Ref.~\cite{CMS-Collaboration:2018ae}.
Results of searches at LHC collected in figure~\ref{fig:Searches-for-fermionic-partners}
and newer results \cite{ATLAS-CONF-2021-040,ATLAS-Collaboration:2022ab}
on the kinds of top partners described so far put bounds on the top
partners mass at around 1.2~TeV.

As mentioned above it is possible to have larger groups and larger
representations in the symmetry breaking pattern. For instance if
the large global symmetry of which breaking the Higgs is a pNGB is
chosen to be $SO(6)$ broken to $SO(5)$ and top quark partners states
are chosen to furnish a $6$-dimensional representation there is one
extra top partners state compared to the case of top quark partners
in the $5$ of $SO(5)$ considered for the minimal model of Ref.~\cite{Agashe:2006aa}.
If we call this new top partner state $\Psi_{1}$, the name signals
the fact that it is a singlet under the remnant $SO(5)$ symmetry,
we can have new signals from its production via QCD interactions and
decay that do not fit into any of the previously considered categories
e.g. 
\[
\Psi_{1}\to th,tZ,t\eta,Wb\;,
\]
where $\eta$ is an extra pNGB that arises due to the larger number
of broken generators in the breaking $SO(6)\to SO(5)\to SO(4)\simeq SU(2)\times SU(2)$
.

In general the extensions of pNGB models can include possible FCNC
of top quarks with new physics states, e.g. Ref.~\cite{Banerjee:2018ab}
has considered decays of the SM top quark that violate flavor
\[
t\to c\eta
\]
as a consequence of underlying flavor-changing dynamics in the top
partners by a coupling $Tc\eta$ which would also yield a new possible
search channel for a top partner $T\to c\eta$. Other exotic possibilities
are covered in the literature, e.g. $T\to tg,t\gamma,$ $X^{5/3}\to t\phi^{+}$
and more exotics ones are presented in Refs.~\cite{1904.05893v1,Kim:2018ad,Bizot:2018aa}
and can in principle lead to new signals for top quark partners.

\section{EFT at current and future colliders\label{sec:EFT}}

The previous sections dealt with explicit models of new physics giving
rise to signals from direct production of particles beyond those of
the Standard Model. As these searches have so far yield no evidence
of new physics a growing interest and motivation have risen for the
description of new physics in Effective Field Theories. The effective
character of these theories is due to the fact that they arise by
the removal of heavy states from a theory more microscopic than the
SM and they lead to a set of BSM interactions, that is usually in
overlap with the set generated by other microscopic theories. Therefore
it has been done a great work in identifying the most general sets
of interactions under given assumptions~\cite{WarsawBasis:2010es,Maltoni:2019aa},
so that new physics studies can be carried in a ``model-independent''
fashion, e.g. searching for very characteristic interactions involving
four top quarks \cite{2010.05915v1,Farina:2018ab,Pomarol:2008nr,Lillie:2008ty}
or other four-fermion operators involving top quarks, or other kinds
of contact interactions independently of their microscopic origin. 

The plus side of the EFT approach is that it is very comprehensive.
The converse of this comprehensiveness is the possible loss of contact
with the microscopic origin of physics beyond the Standard Model which
gives rise to specific patterns and organization principles for the
size of each contact interaction. Thus it is necessary to strike a
balance between a fully general EFT and a ``physically efficacious''
effective theory. Where this balance lies is very much dependent on
the amount of data that one can use in constraining the couplings
of the effective interactions, as well as the theoretical prejudice
on what effects are worth being considered, e.g. pure top sector effects
\cite{2107.13917v1,Hartland:2019aa,Brivio:2019aa,Buckley:2015lr,Buckley:2016aa,Maltoni:2019aa},
or effects involving EW and Higgs physics as well~\cite{2105.00006v1,2012.02779v1}
or exploring flavor changing effects~\cite{Durieux:2014th,Chala:2018aa,Kamenik:2018aa,CMS-Collaboration:2020aa,CMS:2020pnn,ATLAS:2018avw}.

As the effect of BSM contact interactions from the EFT affects precision
measurements of SM processes, this enhanced attention towards signals
of BSM associated to top quarks has produced activity on the improvement
of the description of several processes that are either backgrounds
or serve as SM reference on top of which search for signs of BSM,
e.g. see recent Ref.~\cite{Jezo:2022aa} for four-top production,
recent $ttV$ results discussed in Refs.~\cite{Broggio:2019aa,Bevilacqua:2022aa,Bevilacqua:2018aa},
$tth$ results in Ref.~\cite{Catani:2022aa} and references therein.
For an up to date snapshot of the characterization of the top quark
electroweak interactions and possible BSM in deviations from the SM
we refer the reader to Refs.~\cite{2107.13917v1,Hartland:2019aa,Brivio:2019aa,Buckley:2015lr,Buckley:2016aa}.
The upshot of the work is that present measurements, also thanks
to the availability of differential measurements and trustable computations
in the same phase-space regions, can put bounds on generic new physics
in the top quark sector in the TeV ballpark.

\begin{figure}
\begin{centering}
\includegraphics[clip,width=0.99\textwidth]{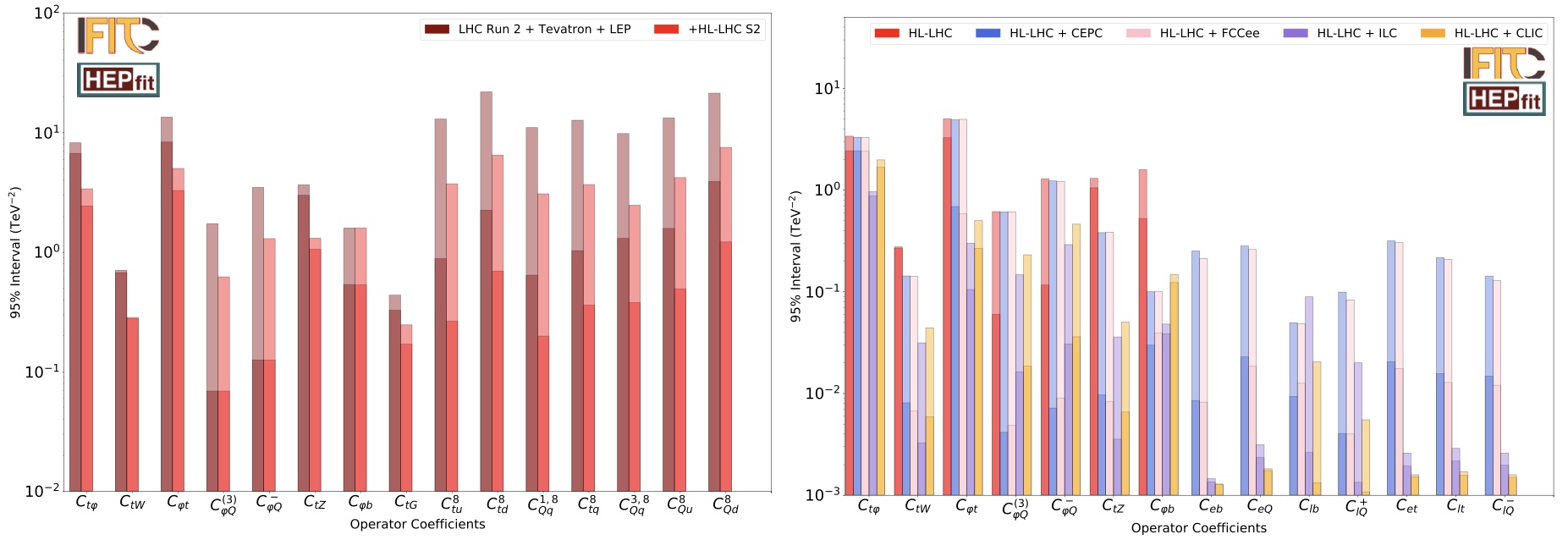}
\par\end{centering}
\caption{\label{fig:EFTcurrent-and-future} Summary from Ref.~\cite{de-Blas:2022aa,Durieux:2022aa}
of the constraints on contact interactions involving the top quark.
The left panel shows the effect of HL-LHC compared to present constraints.
The right panel shows the effect of future $e^{+}e^{-}$ machines.
The taller (and lighted) bars for each case represent the looser bounds
that are obtained when the coupling of interest is bound while the
others are allowed to float, see~\cite{de-Blas:2022aa,Durieux:2022aa}
and references therein for details.}
\end{figure}

The possibility to identify indirect signs of new physics in signatures
related to the top quark has become a commonly used benchmark in the
evaluation of performances of future colliders, especially clean $e^{+}e^{-}$
machines, whose best chance to see new physics in the top sector is
through indirect effects. Works such as \cite{Durieux:2022aa,de-Blas:2022aa,Durieux:2019aa,Durieux:2018ab}
have studied the outcome of analyses to be carried out at future colliders
and the interplay between present and future colliders probes of new
physics in top quark effective field theory. The results are summarized
in Figure~\ref{fig:EFTcurrent-and-future}, which shows the significant
improvement that will be attained by the HL-LHC, especially on single-couplings
effects. The figure also shows the strong tightening of the bounds
with the addition of data from future $e^{+}e^{-}$ data at the $Zh$
threshold, the $t\bar{t}$ threshold and above, which will make the
global EFT constraints particularly robust by the removal of possible
flat directions in couplings-space and providing new data in channels
that can be probed best at clean $e^{+}e^{-}$ machines. 

\section{Top quark and BSM related to Flavor Dynamics or Dark Matter (or both)}

\subsection{Top quark and BSM related to flavor\label{subsec:Top-quark-and-flavor}}

The top quark flavor remains a special one in the SM. Indeed the top
quark is so heavy that one can easily single out the third generation
of quarks as a peculiars source of breaking of the flavor symmetry
\[
G_{F}=U(3)_{q_{L}}\times U(3)_{u_{R}}\times U(3)_{d_{R}}
\]
that the SM would enjoy if all quark masses were zero. A hierarchy
of breaking dominated by the third generation can be accommodated
easily, thanks to the freedom about the possible symmetry breaking
patterns and possible mechanisms for breaking the flavor symmetry
of the SM that one can consider. In addition, this way of organizing
the breaking of flavor symmetry is most compatible with experimental
bounds. In fact, bounds on first and second generation flavor changing
processes are the most tight, whereas there is a relative lack of
constraints on the third generation. If the sole breaking of the symmetry
$G_{F}$ arises from the Yukawa couplings of the SM, or new sources
are aligned with the Yukawa matrices, the breaking is said to comply
with ``Minimal Flavor Violation''~(MFV)~\cite{DAmbrosio:2002aa,Buras:2001aa,Buras:2003aa}.
In this setting the bounds from flavor observables are most easily
accommodated, but it is not the only possibility to comply with observations.
The fact that the top quark Yukawa coupling is a possible large source
of flavor symmetry breaking motivates to consider BSM related to the
top flavor, but this conclusion holds also in other settings.

A classification of possible states that can couple to quark bilinears
charged under the flavor symmetry, e.g. a new scalar coupled as $\phi_{tu}tu$,
has proven useful in the past to assess the possibility of flavorful
signs of new physics. For a recent listing of the possible states
one can read tables on Ref.~\cite{Grinstein:2011kl}. From a phenomenological
point of view these models give rise to transitions in four-quark
scatterings that do not conserve the flavor charge. For instance the
scattering 
\[
uu\to tt
\]
can arise via a $t$-channel exchange of a flavored boson. This can
alter the kinematic of top quark production as well as the net charge
of the top quark sample at hadron colliders. Indeed new flavorful
boson of this kind were advocated in response to TeVatron experiments
claiming disagreements between the SM predictions and measured top
quark properties, such as the forward-backward asymmetry in the production
of top quarks~\cite{Aaltonen:2011aa,Abazov:2011af,CDF-Collaboration:2012aa}.
In addition these new flavored states coupled to the top quark can
give rise to transitions 
\[
f\bar{f}\to t\phi_{tj}u_{j}\,,
\]
that can be observed quite easily at $e^{+}e^{-}$ colliders in multi-jet
final states, the detailed final state depending on the model-dependent
decay of the flavored state $\phi$. 

The possibility that a flavored state connected to the top quark might
be among the lightest new states from the new physics sector has appeared
also in models of gauged flavor symmetries. In these models the flavor
symmetry $G_{F}$ is gauged, as to not have to deal with unobserved
massless Goldstone bosons. For instance Refs.~\cite{Grinstein:2010uq,Berezhiani:1983hm}
have proposed a new set of states that would have the notable property
to make the $G_{F}$ gauging free from triangular anomalies by the
addition of vector-like new quarks. In this kind of models the new
quarks are charged under the SM flavor symmetry and can be arranged
as to have top-flavor new states to be the lightest ones. Indeed in
these models the masses of the the SM quarks would be explained by
a see-saw-like mechanism in which the lightest SM fermions are mixed
with a very heavy new state, whereas the heaviest SM states are mixed
with the lightest of the new physics states. In this case the SM top
quark would be the state coupled to the lightest of the new physics
states, named $t'$, possibly accompanied by a partner state for the
bottom quark, named $b'$. Remarkably this type of model gives phenomenological
signatures very similar to those of top partners states of composite
and little Higgs, e.g. the partner states can be produced by strong
interactions and decay as 
\[
b'\to bh,bZ,tW
\]
 and 
\[
t'\to th,tZ,bW\,.
\]

These ideas also lend themselves to be paired with supersymmetry.
Although supersymmetry is not necessary for the idea of gauged flavor
symmetries in general, these models can provide a setup to originate
R-parity breaking with an underlying structure for the flavor structure
of the RPV couplings~\cite{Krnjaic:2012rw,Franceschini:2013ne},
that for instance would motivate 
\[
\tilde{t}\to bs
\]
 as the main channel to search RPV stops~\cite{Franceschini:2012fu}. 

\begin{figure}
\centering{}\includegraphics[width=0.67\paperwidth]{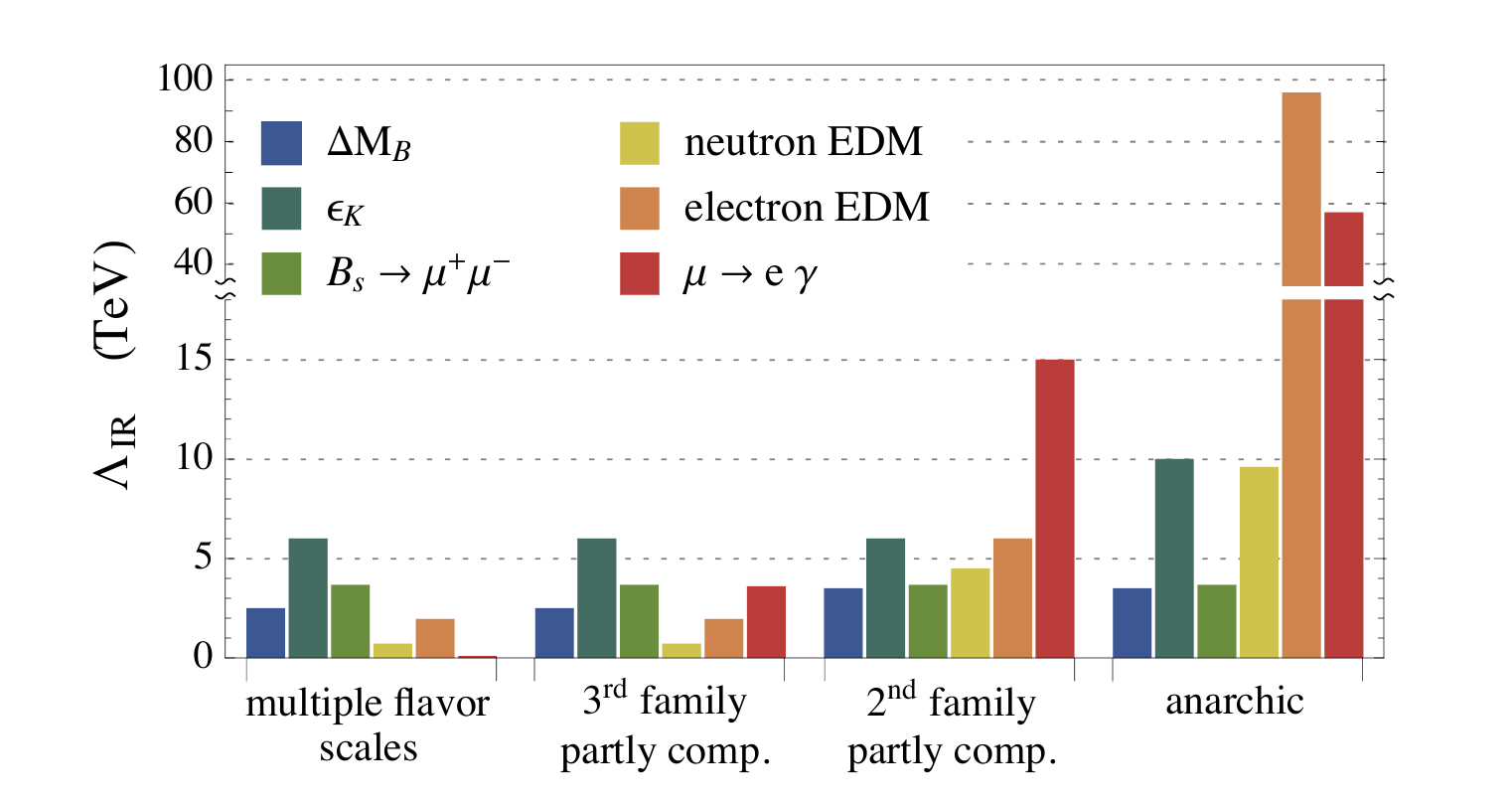}\caption{\label{fig:Lower-bound-on-CH-flavor}Lower bound on the scale of new
physics related to the SM fermion mass generation in a composite Higgs
scenario~\cite{Panico:2016kx} under different assumptions on the
compositeness of SM fermions.}
\end{figure}

A solution with a hierarchy of flavored new physics scales inverted
with respect to that of the SM quarks has been proposed also for composite
Higgs models~\cite{Panico:2016kx,1206.4701v3,Cacciapaglia:2015aa,Matsedonskyi:2015aa},
which would otherwise suffer from severe bounds from high-$p_{T}$
and flavor observables (see e.g.~\cite{Keren-Zur:2013qe,Domenech:2012qy,Redi:2013th}),
even in presence of some degree of model building \cite{Redi:2012aa,Redi:2011aa,Delaunay:2014aa}
aimed at keeping all the new physics at a common low-scale and still
survive flavor tests thanks to a friendly, possibly MFV-like structure,
of the flavor origin in the microscopic completion of the composite
Higgs model. As it can be appreciated in Fig.~\ref{fig:Lower-bound-on-CH-flavor}
the top quark sector emerges still as a less constrained one and further
motivates to consider BSM physics related to the top quark, and possibly
exclusively to the top quark or to the third generation of SM fermions.

Observables of interests include indirect probes such as electric
dipoles moments (see e.g. \cite{1712.06337v2}), meson oscillations
and decays, and in principle rare $Z$ and Higgs bosons flavor-violating
decays which usually receive important contributions from the top
quark sector~\cite{Panico:2016kx}. In addition, it is possible to
have phenomena more directly related to the top quark such as 
\[
t\to cV,
\]
where $V=\gamma,Z,g$\cite{Agashe:2007aa,hep-ph/0408134v1} and deviations
from $V_{tb}=1$ in the CKM matrix~\cite{Matsedonskyi:2016aa,Grojean:2013yq,del-Aguila:2000aa,Aguilar-Saavedra:2013aa}. 

\subsection{Flavored dark matter models\label{subsec:Flavored-dark-matter}}

Given the strength of the bounds from direct searches of dark matter
scattering on heavy nuclei it has become interesting to consider dark
matter models in which the flavor of SM quarks and leptons plays a
role, as the strongest bounds hinge on effective couplings of the
dark matter to first and, to a slightly lesser extent, to second generation
quarks and gluons. 

Rather interestingly the flavor puzzle of the SM comes equipped with
a symmetry, which, though not exact, can be used to stabilize the
dark matter if it is broken according to Minimal Flavor Violation~\cite{1105.1781v2,1308.0584v1}
and even with more general patterns of flavor symmetry and its breaking~\cite{1405.6709v2}.
As a dark matter coupling sensitive to flavor could mediate flavor
changing transitions the option of the MFV structure, or slight departures
from it, has been so far been a main route in model building aimed
at removing possible tensions with flavor observables.

Among the possible flavor structures that the Dark Matter and the
SM can fields can be cast in, for our work here we focus on the possibility
that the top quark flavor has a special role. Explicit models have
appeared in the context of possible explanations of the CDF $A_{FB}$
anomaly~\cite{Aaltonen:2011aa,Abazov:2011af,CDF-Collaboration:2012aa},
e.g. see the model built in Ref.~\cite{Kumar:2013ab}, but the idea
stands out on itself even without anomalies in top quark physics.
Indeed if one considers that the complexity of the SM may be replicated
in the sector of dark matter it is natural to consider multiple species
of dark matter, that are ``flavors'' of dark matter~\cite{Kile:2011aa,Kamenik:2011aa,Agrawal:2012ab}.
These flavors can be separated from our own SM flavors or can be related
to our species of fermions. In case some relation exists between flavors
of the SM and of the dark sector the possibility that the top quark
flavored dark matter is the lightest state is at least as probable
as any other flavor assumption. For example, when Minimal Flavor Violation
is advocated one can explicitly write a mass term for the dark-flavor
fermion multiplet $\chi$ which in general has the form
\[
\bar{\chi}\left(m_{0}+\varUpsilon(YY)\right)\chi\,,
\]
where $\varUpsilon$ is a function of combinations of the Yukawa matrices
of the SM that form singlets under the flavor group that is dominated
by the piece proportional to $Y_{u}^{\dagger}Y_{u}$, hence the top
quark flavor tends to be special just from the principle of MFV itself.
In a concrete case we can have interactions of SM fermions $u_{R}^{(i)}$
and mass terms for the dark matter flavor multiplet $\chi$
\begin{equation}
\phi\bar{\chi}\left(g_{0}+g_{1}Y_{u}^{\dagger}Y_{u}\right)u_{R}^{(i)}+h.c.+\bar{\chi}\left(m_{0}+m_{1}Y_{u}^{\dagger}Y_{u}+...\right)\chi\,,\label{eq:MFVtopDM}
\end{equation}
where $\phi$ is a suitable representation of $G_{SM}\otimes G_{F}$.
In Ref.~\cite{Kumar:2013ab} for instance $\phi\sim(\mathbf{3},\mathbf{1},2/3)_{SM}\otimes(\mathbf{1},\mathbf{1},\mathbf{1})_{F}$,
$\chi\sim(\mathbf{1},\mathbf{1},0)_{SM}\otimes(\mathbf{1},\mathbf{3},\mathbf{1})_{F}$
and the Yukawa matrices, as in general in MFV, transform as spurions
$Y_{u}\sim(\mathbf{3},\mathbf{\bar{3}},1)_{F}$ and $Y_{d}\sim(\mathbf{3},1,\mathbf{\bar{3}})_{F}$.
We see that it is possible to pick $m_{1}$ as to partly cancel the
flavor universal $m_{0}$ term, making $\chi_{t}$ the lightest particle
of the $\chi$ multiplet while retaining full freedom to pick the
combinations of $g_{0}$ and $g_{1}$ that corresponds to the couplings
of the mass eigenstates $\chi_{i}$. 

In absence of a field $\phi$ one can imagine contact operators to
couple the Dark Matter and the SM flavors $i$ and $j$, e.g. operators
of the type 
\begin{equation}
\left(\bar{\chi}\Gamma_{S}\chi\right)\left(\bar{\psi}^{(i)}\Gamma_{S}\psi^{(j)}\right)\label{eq:4fermionsDMSM}
\end{equation}
for some Lorentz structure $\Gamma_{S}$ have been considered as low
energy remnants of flavored gauge bosons~\cite{Kile:2011aa} or other
heavy scalar and fermion states charged under a MFV-broken flavor
symmetry or in a horizontal symmetry model~\cite{Kamenik:2011aa}.
Operators involving the SM Higgs boson, e.g.
\[
\left(\bar{Q}\chi\right)\left(\chi^{*}Hu\right)
\]
have also been considered in \cite{1105.1781v2} for a scalar $\chi\sim(\mathbf{1},\mathbf{1},0)_{SM}\otimes(\mathbf{3},\mathbf{1},\mathbf{1})_{F}$.
A variation of the model of Ref.~\cite{Kile:2011aa} could lead to
top quark flavor being singled out, the other referred works already
consider the third generation, hence the top quark and/or the bottom
quark, as special due to either the MFV structure or as a result of
the horizontal symmetry. 

The phenomenology of top flavored dark matter is very rich as it comprises
both possible signals in dark matter searches and in precision flavor
observables as well as in high energy collider searches. Flavor observables
put in general stringent bounds on flavored dark matter models, the
case of top-flavored dark matter being significantly less constrained
due to majority of data belonging to $u,d,s,c,b$ quark systems. Dark
matter direct detection is also in general suppressed because nucleons
involved in dark matter scattering do not contain top flavor, hence
the interactions are usually originated at loop level or via breaking
of the flavor alignments, i.e. the dark matter interacts almost exclusively
with top quark flavor, but it may have a small, though not completely
negligible coupling to light flavors. The existence of such coupling
depends on the model. A specific analysis for a case in which only
top quark flavor interacts with the DM in the model eq.(\ref{eq:MFVtopDM})
is presented in Ref.~\cite{Kilic:2015eu} for both dark matter direct
detection and collider prospects in a MFV scenario. The annihilation
rate for the thermal freeze-out is set by the scattering
\begin{equation}
\chi\chi\to tt\label{eq:chichi2tt}
\end{equation}
mediated by a mediator $\phi$ (other scatterings are discussed in
detail for instance in \cite{Garny:2018aa}). In this specific case
the direct detection scattering on nucleons
\[
\chi\mathcal{N}\to\chi\mathcal{N}
\]
is mediated by a loop induced couplings of $Z,\gamma$ to $\chi$
from a bubble loop of $t$ and $\phi$ from eq.(\ref{eq:MFVtopDM}).
Despite the smallness of these couplings the reach of current and
future large exposure experiments, e.g. see \cite{2203.02309v1},
could probe such low level of scattering rates for exposure around
1 ton year, that means the model can be tested with presently available
data~\cite{Aprile:2018aa}.

A more recent analysis~\cite{1702.08457v2} considered flavor, direct
dark matter detection and collider searches for a model featuring
a top-flavored dark matter $\chi$ and a new state $\phi$. In this
work a ``Dark Minimal Flavor Violation'' flavor structure that extends
MFV, but can recover it as a limit, is considered and allows for a
more generic structure in flavor space for the vertex
\[
\lambda_{ij}\bar{u}_{R}^{(i)}\phi\chi+h.c.\,.
\]
In this context it is possible to delay the observation of $\chi$
in direct detection experiments, as new contributions to the direct
detection rate appear compared to the MFV case and it is possible
to arrange for cancellations among scattering amplitudes. It remains
an open questions if it is going to be possible to claim an observation
in spite of the so-called ``neutrino fog'' that future Xenon experiments~\cite{2203.02309v1}
face when probing rates so small that neutrinos from the Sun, supernovae
and other natural sources are expected to contribute an event rate
comparable or larger than that of the dark matter.

In principle it is possible to have $m_{\chi}<m_{t}$ so that the
thermal freeze-out is controlled by other processes than the simple
tree-level exchange of eq.(\ref{eq:chichi2tt}). Reference \cite{1702.08457v2}
experimented with this possibility in Dark Minimal Flavor Violations,
but it appears in tension with the direct detection experiments. This
conclusion concurs with what can be extrapolated from the earlier
MFV analysis of \cite{Kumar:2013ab}.

The search for models with mediators, that are colored in all models
considered so far, can be carried out very effectively at hadron colliders
searching for signals
\[
pp\to\phi\phi\to t\chi t\chi\,,
\]
that very much resemble the search for supersymmetric top partners.
Depending on the model there can be more general combinations of flavors
of quarks
\[
pp\to\phi\phi\to q_{j}\chi q_{i}\chi\,.
\]
Therefore it is in general useful to consider the whole list of squark
searches to put bounds on this type of models. References \cite{1702.08457v2,Blanke:2020bsf}
reports bounds in the TeV ballpark which inherit the strengths and
weaknesses discussed for the search of supersymmetric quark partners.

Other possible signals at hadron collider are the 
\[
pp\to t\chi\chi
\]
scattering, which can arise from interactions such as eq.(\ref{eq:4fermionsDMSM}),
studied in \cite{Kamenik:2011aa}, or associated production $\phi\chi$,
followed by $\phi\to t\chi$ studied for instance in \cite{Blanke:2020bsf}.

It is also possible to consider models that go beyond what we have
considered here starting from the notable feature that MFV and some
extensions may render the DM stable. In a model of such ``top-philic''
dark matter model on can have \cite{Andrea:2011aa} scalars that couple
to $t\chi$ as well as to light quark bilinears, e.g. from RPV supersymmetry,
so that they mediate scatterings of the type
\[
q_{i}\bar{q}_{j}\to S_{ij}\to t\chi\,.
\]
Other potentially interesting signals possible flavored gauge bosons
with couplings $\rho_{ij}q_{i}q_{j}$ can appear, replacing $S_{ij}$
with $\rho_{ij}$ in the above process. Further signals in this type
of models arise, e.g. 
\[
q_{i}g\to t\rho_{ti}
\]
possibly followed by $\rho\to\chi t$, and similarly for $S$. A model
with a flavored gauge boson has been studied in \cite{DHondt:2015yg}
with the goal of pinning down the flavor of light quark that interacts
with the top quark and the dark matter leveraging charm-tagging and
lepton charge asymmetry at the LHC.

Though many general issues follow the same path for scalar and fermionic
dark matter it is worth mentioning that references~\cite{Colucci:2018aa,1709.00697v5}
contain a full study of the case in which the partner and the dark
matter are a fermion and a scalar, respectively, at the converse of
most of what we discussed above. Further studies of top and dark matter
related matters can be found in the context of simplified models building
\cite{Garny:2018aa,1404.1918v2,Cheung:2010ab}.

\section{Conclusions\label{sec:Conclusions}}

The connection between new physics and the top quark sector is well
established and has lead to a large amount of model building and phenomenological
studies. Here we have presented supersymmetric top partners, motivated
by supersymmetry as the symmetry that stabilizes the weak scale, and
top partners states motivated by the possible compositeness and pseudo-Nambu-Goldstone
boson nature of the Higgs boson. The phenomenological relevance of
these incarnations of ``BSM in the top quark sector'' is tightly
tied to the motivations of the models to which the top partners states
belong. As the models in question are themselves in a ``critical''
phase at the moment, so is the situation for this type of new physics
in the top quark sector. We say this in the sense that on one hand
we have reached a point at which the expectation was to have already
discovered signs of new physics, especially in the top quark sector
in the mass range explored by current experiments, hence we should
start to dismiss these ideas, while on the other hand we are still
largely convinced of the validity of the arguments that lead to the
formulation of these models. Furthermore no serious alternatives have
appeared in the model building landscape and we still have plenty
of evidence for the existence of physics beyond the Standard Model.
Thus one can be lead to reconsider if the entire motivational construction
for these models was somewhat wrong or at least biased towards a ``close-by''
and experimentally friendly solution. 

The way out of this crisis, in absence of experimental results changing
the situation, is for everyone to decide. A possibility is to conclude
that we need to update our beliefs about ``where'' \cite{Arkani-Hamed:2012rw}
new physics can appear in the top quark sector and more in general
in going beyond the SM. In this sense top partner searches are a gauge
of our progress on testing well established ideas on new physics. 

It should be remarked that the top quark sector remains central also
in the formulation of new physics models that try alternatives to
the more well established ideas, see e.g. Refs.~\cite{Bally:2022aa,KimHD22017}
on possible ways the top quark can lead the way to construct new physics
models of a somewhat different kind that the two mainstream ideas
discussed here.

Given the absence of clear signs and directions in model building
into which entrust our hopes for new physics we have discussed the
power of general effective field theory analyses that can be used
to search for new physics in precise SM measurements. These tools
have become the weapon of choice in a post-LHC epoch for the so-called
model-independent search of new physics. We have presented the power
of current LHC and future HL-LHC analyses to see deviations from the
SM due to top quark interactions. Overall the LHC has a chance to
see deviation in some more friendly observables for a new physics
scale in the TeV range. In order to secure this result and avoid possible
blind-spots a new particle accelerator is needed, a most popular option
being an $e^{+}e^{-}$ capable of operating at or above the $t\bar{t}$
threshold with the luminosity to produce around $10^{6}$ top quark
pairs.

Other great mysteries beyond the origin of the electroweak scale remain
unsolved in the Standard Model. We have looked at possible solutions
of the flavor puzzle in which the top quark flavor plays a special
role. The phenomenology of models with lowest lying new physics states
charged under top flavor has some similarity with that of top quark
partners at colliders, but there is also the possibility to generate
observable flavor violations as further distinctive experimental signatures.

We have examined the possibility that the top quark may be a key to
solve the mystery of dark matter of the Universe. We have presented
scenarios in which the dark matter interacts predominantly or exclusively
with the top quark flavor, possibly ascribing the stability of the
dark matter to the same flavor structure that makes the top quark
flavor special among the SM flavors. Such possibility appears very
well motivated as a way to reduce otherwise intolerably large couplings
of dark matter with lighter generations and explain the stability
of dark matter. The flavor dependence of the couplings has motivated
efforts to build models for the realization of this idea in a coherent,
though maybe still effective, theory of favor of which we have presented
a few instances. We remarked how in these scenarios the dark matter
phenomenology is quite different from other types of thermal dark
matter and we have summarized dedicated analyses that have been carried
out to identify the relevant bounds and constraints. The upshot is
that idea can be broadly tested with current and future direct detection
dark matter experiments. At the same time the new states associated
with the dark matter may be observed on-shell at colliders, which
can in principle also probe contact interactions that originate from
off-shell states associated with the dark matter. Low energy flavor
observables can also help to restrict the range of possible models
of flavored dark matter leading to significant constraints both on
MFV and non-MFV scenarios when a thermal relic abundance and a significant
suppression of spin-dependent and spin-independent direct detection
rates are required.

\subsection*{Acknowledgments}
It is a pleasure to thank Kaustubh Agashe for discussions on top quark partners in composite Higgs models. 

\providecommand{\href}[2]{#2}\begingroup\raggedright\endgroup


\end{document}